\documentclass[pss,fleqn]{w-art}
\usepackage{times}
\usepackage{w-thm}
\usepackage[]{graphicx}
\newcommand{\up}{\uparrow}
\newcommand{\dn}{\downarrow}

\begin{document}

%%    The information for the title page will be placed between
%%    \begin{document} and \maketitle. The order of most entries
%%    is determined by the class file and can not be changed by
%%    rearranging them. The maketitle command follows after the
%%    absract.
%%
%%    Most of the following commands will be completed by the publisher.
%%
\DOIsuffix{theDOIsuffix}
%%
%% issueinfo for header and copyright line
\Volume{XX}
\Issue{1}
\Copyrightissue{01}
\Month{01}
\Year{2004}
%%
%%    First and last pagenumber of the article. If the option
%%    'autolastpage' is set (default) the second argument may be left empty.
\pagespan{1}{}
%%
%%    Dates will be filled in by the publisher. The 'reviseddate' and
%%    'dateposted' (Published online) entry may be left empty.
%%%\Receiveddate{\sf zzz} \Reviseddate{\sf zzz} \Accepteddate{\sf zzz} \Dateposted{\sf zzz}
%%
%%    Give a maximum of six PACS code in numerical order.
%[{\bf You must insert up to 6 relevant codes, see} {\sf www.aip.org/pacs.}]

%% \pretitle{Editor's Choice}

%% We have a short and a long form for the title. The short form
%% (optional argument) goes into the running head.

\title{Bosonization of dimerized Hubbard chains}

%% Please do not enter footnotes or \inst{}-notes into the optional
%% argument of the author command. The optional argument will go into
%% the header.  If there is only one address the marker \inst{x} may be
%% omitted.

%% Information for the first author.
\author{C.\ Mocanu\footnote{Corresponding author E-mail: {\sf carmen.mocanu@physik.uni-augsburg.de}}, M.\ Dzierzawa, P.\ Schwab, and U.\ Eckern}
\address{Institut f\"ur Physik, Universit\"at Augsburg, 86135 Augsburg, Germany}\

\begin{abstract}
The role of Klein factors is investigated for the bosonized Hamiltonian
of the dimerized Hubbard model. Contrary to previous approaches
we take into account their number changing property, i.e.\
we do not replace them by Majorana fermions.
We show how to treat Klein factors
in the framework of the self-consistent harmonic approximation,
both for finite systems and in the thermodynamic limit.

\end{abstract}

%% maketitle must follow the abstract.
\maketitle                   % Produces the title.

%% If there is not enough space inside the running head
%% for all authors including the title you may provide
%% the leftmark in one of the following three forms:

%% \renewcommand{\leftmark}
%% {First Author: A Short Title}

%% \renewcommand{\leftmark}
%% {First Author and Second Author: A Short Title}

\renewcommand{\leftmark}
{C. Mocanu et al.: Bosonization of dimerized Hubbard chains}

% 71.10.-w Theories and models of many-electron systems
% 71.10.Fd Lattice fermion models (Hubbard model, etc.)
% 71.10.Pm Fermions in reduced dimensions (anyons, composite fermions,
%         Luttinger liquid, etc.) (for anyon mechanism in superconductors, see
%         74.20.Mn)
% 71.30.+h Metal-insulator transitions and other electronic transitions
%\maketitle

\section{Introduction}
The foundations of bosonization were laid more than
50 years ago in a seminal paper by Tomonaga \cite{Tomonaga50}.
During the following decades the method was worked out
and successfully applied to one-dimensional
electron and spin systems \cite{Haldane81,Delft98,Schulz00}.
Despite its long history there are still some subtle points in
the bosonization formalism
which are not taken into consideration in the majority of the literature.
One of these is the proper treatment of the so-called
Klein factors which have to be introduced in order
to preserve the anticommuting property of the fermionic fields
during the bosonization procedure.
The role of the Klein factors deserves particular attention
when nonlinear perturbations arising e.g.\ from impurity scattering
or lattice modulations are to be considered in finite systems.

In this paper we demonstrate how to handle the Klein factors 
in a systematic way, both in the thermodynamic limit and for finite systems.
As a prototypical model we study the
one-dimensional dimerized Hubbard model
where the hopping is periodically modulated due to the Peierls
distortion of the lattice.
We extend the self-consistent harmonic approximation
\cite{Coleman75,Fukuyama85,Gogolin93}
by treating bosonic fields and Klein factors on equal footing
\cite{Mocanu04}.
As an application we use the formalism to calculate
spin and charge gaps of this model.

\section{Bosonization and Klein factors}

We consider the dimerized Hubbard model
%%%%%%%%%%%%%%%%%%%%%%%%%%%%%%%%%%%%%%%%%%%%%%%%%%%%%%%%%%%%%%%
\begin{equation}
\label{PHmodel}
H = -t \sum_{i,\sigma} (1 + (-1)^i u)(c_{i\sigma}^+ c_{i+1,\sigma}^{}
+ c_{i+1,\sigma}^+ c_{i\sigma}^{})
+ U \sum_{i} n_{i\up} n_{i\dn}
\end{equation}
%%%%%%%%%%%%%%%%%%%%%%%%%%%%%%%%%%%%%%%%%%%%%%%%%%%%%%%%%%%%%%%
which differs from the ordinary one-dimensional Hubbard model by a periodic modulation
of the hopping
described by the dimerization parameter $u$. This modulation is relevant in one
dimension due to the coupling between the lattice and the electronic degrees of
freedom.
A finite $u$ corresponds to a periodic lattice distortion known as Peierls instability.
We study the case of half-filling, and have chosen the modulation accordingly to
be of the form $(-1)^i u$.
The quantity $u$ is considered to be a parameter of the model.
$c_{i\sigma}^+$ creates an electron with spin direction $\sigma = \up,\dn$ at site
$i$, and $t$ and $U$ are the hopping matrix element and the on-site Hubbard interaction,
respectively.

In the following we list the main steps which have to be performed in order
to bosonize the Hamiltonian (\ref{PHmodel}). For more details we refer the
reader to the reviews \cite{Delft98,Schulz00,Fukuyama85}.
First we represent $H$ in momentum space:
%%%%%%%%%%%%%%%%%%%%%%%%%%%%%%%%%%%%%%%%%%%%%%%%%%%%%%%%%%%%%%%
\begin{equation}
\label{PHmodelk}
H = \sum_{k,\sigma} \epsilon_k c_{k\sigma}^+ c_{k\sigma}^{}
+ t u \sum_{k,\sigma} ({\rm e}^{ik} c_{k\sigma}^+ c_{k+\pi,\sigma}^{} + {\rm h.c.})
+ \frac{U}{N} \sum_{k,k',q} c_{k+q,\up}^+ c_{k\up}^{} c_{k'-q,\dn}^+ c_{k'\dn}^{}
\end{equation}
%%%%%%%%%%%%%%%%%%%%%%%%%%%%%%%%%%%%%%%%%%%%%%%%%%%%%%%%%%%%%%%
where $\epsilon_k = - 2t \cos k$, and $N$ is the number of lattice sites.
We linearize the spectrum around
the two Fermi points $\pm k_F$,\ $k_F = \pi/2$, and introduce left and right moving
fermions labeled by $L$ and $R$, respectively. This allows us to sort
the various scattering processes according to their
initial and final states (``g-ology").
Then we define bosonic operators $b_{q\alpha}^+$
which are related to fermionic particle-hole excitations via
[$\Psi_{L/R \sigma}( k ) = c_{\mp(k_F + k), \sigma}$ etc.; $c_{k \sigma}^{}$, $ c_{k \sigma}^{+}$ are the standard Fermion 
operators]
%%%%%%%%%%%%%%%%%%%%%%%%%%%%%%%%%%%%%%%%%%%%%%%%%%%%%%%%%%%%%%%
\begin{equation}
\label{bq}
b_{q\alpha}^+ = - \frac{i}{\sqrt{n_q}} \sum_{k} \Psi^+_\alpha({k+q)} \Psi_\alpha^{}({k }) \quad ( q > 0 )
\end{equation}
%%%%%%%%%%%%%%%%%%%%%%%%%%%%%%%%%%%%%%%%%%%%%%%%%%%%%%%%%%%%%%%
where $q = (2\pi/N) n_q$ and $\alpha = R\up,L\up,R\dn,L\dn$. Returning to real space
we define the bosonic fields
%%%%%%%%%%%%%%%%%%%%%%%%%%%%%%%%%%%%%%%%%%%%%%%%%%%%%%%%%%%%%%%
\begin{equation}
\label{varphi}
\varphi_{L \sigma }(x) =   \sum_{q > 0} \frac{1}{\sqrt{n_q}} b_{q L \sigma} {\rm e}^{-i qx - aq/2} \; , \;
\varphi_{R \sigma }(x) = - \sum_{q > 0} \frac{1}{\sqrt{n_q}} b_{q R \sigma} {\rm e}^{i qx - aq/2} \; ,
\;\; a\rightarrow 0
\end{equation}
%%%%%%%%%%%%%%%%%%%%%%%%%%%%%%%%%%%%%%%%%%%%%%%%%%%%%%%%%%%%%%%
which are related to the fermionic field operators
via the bosonization identities
%%%%%%%%%%%%%%%%%%%%%%%%%%%%%%%%%%%%%%%%%%%%%%%%%%%%%%%%%%%%%%%
\begin{eqnarray}
\label{bosident}
\Psi_{L \sigma} (x) & = & \frac{1}{\sqrt{L}} F_{L\sigma}
{\rm e}^{-i\varphi_{L\sigma }^+(x)} {\rm e}^{-i\varphi_{L \sigma}(x)}
{\rm e}^{-2\pi i N_{L\sigma} x/L} \\
 & = &
\frac{1}{\sqrt{2 \pi a}} F_{L \sigma} 
{\rm e}^{-i(\varphi_{L\sigma}^+(x) + \varphi_{L \sigma}(x))}
{\rm e}^{-2\pi i N_{L \sigma}x/L} \\
\Psi_{R \sigma} (x) & = &
\frac{1}{\sqrt{2 \pi a}} F_{R \sigma}
{\rm e}^{+i(\varphi_{R\sigma}^+(x) + \varphi_{R \sigma}(x))}
{\rm e}^{+2\pi i N_{R \sigma}x/L}
\end{eqnarray}
%%%%%%%%%%%%%%%%%%%%%%%%%%%%%%%%%%%%%%%%%%%%%%%%%%%%%%%%%%%%%%%
where $N_\alpha$ counts the particle number with respect to
the filled Fermi sea, and $L$ is the length of the 
system.\footnote{We have chosen the lattice constant to be unity, i.e.\ $L=N$. Nevertheless
we retain $L$ and $N$ for easy reference.}
The Klein factors $F_\alpha^+$ ($F_\alpha$) are unitary
operators that commute with the bosonic fields. They change the number
of the fermion species $\alpha$ by $\pm 1$, a change which cannot be achieved
by any combination of the bosonic field operators. In order to ensure
the correct anticommutation relations for $\Psi_\alpha(x)$, the
Klein factors have to fulfill the following relations:
%%%%%%%%%%%%%%%%%%%%%%%%%%%%%%%%%%%%%%%%%%%%%%%%%%%%%%%%%%%%%%%
\begin{equation}
\{F_{\alpha},F_{\beta}\} = \{F_{\alpha}^+,F_{\beta}^+\} = 0, \;\;\;\;\;\;
\{F_{\alpha}^+,F_{\beta}\} = 2 \delta_{\alpha\beta}, \;\;\;\;\;\;
[N_{\alpha},F_{\beta}] = - \delta_{\alpha \beta} F_{\beta}\;
.\end{equation}
%%%%%%%%%%%%%%%%%%%%%%%%%%%%%%%%%%%%%%%%%%%%%%%%%%%%%%%%%%%%%%%
Here $[.,.]$ denotes the commutator, and $\{.,.\}$ the anticommutator.
In a last step one combines
$\varphi_\alpha$ and $\varphi_\alpha^+$ to
new fields $\phi_{c,s}$ and $\theta_{c,s}$,
%%%%%%%%%%%%%%%%%%%%%%%%%%%%%%%%%%%%%%%%%%%%%%%%%%%%%%%%%%%%%%
\begin{eqnarray}
\phi_{c,s} &= & \frac{1}{2\sqrt{2}}(\varphi_{L \uparrow} \pm \varphi_{L \downarrow } 
                              + \varphi_{R \uparrow} \pm \varphi_{R \downarrow}+ {\rm h.c.} ) \\
\theta_{c,s} &= & \frac{1}{2\sqrt{2}}(\varphi_{L \uparrow} \pm \varphi_{L \downarrow } 
                              - \varphi_{R \uparrow} \mp \varphi_{R \downarrow}+ {\rm h.c.} ) 
\end{eqnarray}
%%%%%%%%%%%%%%%%%%%%%%%%%%%%%%%%%%%%%%%%%%%%%%%%%%%%%%%%%%%%%%
and
introduces $N_{c,s} = N_\up \pm N_\dn$ and $J_{c,s} = J_\up \pm J_\dn$
where $N_\sigma = N_{L\sigma} + N_{R\sigma}$ and $J_\sigma = N_{L\sigma} - N_{R\sigma}$.
Combining everything we obtain
%%%%%%%%%%%%%%%%%%%%%%%%%%%%%%%%%%%%%%%%%%%%%%%%%%%%%%%%%%%%%%%
\begin{equation}
\label{hamilt}
H = H_0 + H_1 + H_2 + H_3
\end{equation}
%%%%%%%%%%%%%%%%%%%%%%%%%%%%%%%%%%%%%%%%%%%%%%%%%%%%%%%%%%%%%%%
where $H_0$ is the Luttinger Hamiltonian:
%%%%%%%%%%%%%%%%%%%%%%%%%%%%%%%%%%%%%%%%%%%%%%%%%%%%%%%%%%%%%%%
\begin{equation}
\label{Luttinger}
H_{0}  = \sum_{\alpha=c,s}\int_0^L
    \frac{dx}{2\pi} : \left\{\frac{v_\alpha}{g_\alpha}(\partial_x\phi_\alpha)^2 +
      v_\alpha g_\alpha (\partial_x\theta_\alpha)^2 \right\} :
      + \frac{\pi}{4L}\sum_{\alpha=c,s}\left\{\frac{v_\alpha}{g_\alpha}N^2_\alpha +
        v_\alpha g_\alpha J^2_\alpha\right\}
    \;\; ;
\end{equation}
%%%%%%%%%%%%%%%%%%%%%%%%%%%%%%%%%%%%%%%%%%%%%%%%%%%%%%%%%%%%%%%
$H_1$ the Umklapp contribution:
%%%%%%%%%%%%%%%%%%%%%%%%%%%%%%%%%%%%%%%%%%%%%%%%%%%%%%%%%%%%%%%
\begin{equation}
\label{Umklapp}
H_1  = \tilde U \int_0^L dx \;
   \{F^+_{R\up}F^+_{R\dn}F_{L\dn}F_{L\up}
    {\rm e}^{-i 2\sqrt{2}\phi_c} + {\rm h.c.}\} \; ,
    \;\;\;\;\;\;\; \tilde U = \frac{U L}{(2\pi a)^2 N}
    \;\; ;
\end{equation}
%%%%%%%%%%%%%%%%%%%%%%%%%%%%%%%%%%%%%%%%%%%%%%%%%%%%%%%%%%%%%%%
$H_2$ the backscattering contribution:
%%%%%%%%%%%%%%%%%%%%%%%%%%%%%%%%%%%%%%%%%%%%%%%%%%%%%%%%%%%%%%%
\begin{equation}
\label{Backscattering}
H_2 = \tilde U \int_0^L dx \;
   \{ F^+_{R\up}F^+_{L\dn}F_{R\dn}F_{L\up}
    {\rm e}^{-i 2\sqrt{2}\phi_s} + {\rm h.c.} \}
    \;\; ;
\end{equation}
%%%%%%%%%%%%%%%%%%%%%%%%%%%%%%%%%%%%%%%%%%%%%%%%%%%%%%%%%%%%%%%
and $H_3$ the dimerization contribution ($\tilde u = tu / \pi a$):
%%%%%%%%%%%%%%%%%%%%%%%%%%%%%%%%%%%%%%%%%%%%%%%%%%%%%%%%%%%%%%%
\begin{equation}
\label{Peierls}
H_3 =
   \tilde u \int_0^L dx \;  \{i F^+_{R\up}F_{L\up}
   {\rm e}^{-i\sqrt{2}(\phi_c+\phi_s)}
   +                                  i F^+_{R\dn}F_{L\dn}
   {\rm e}^{-i\sqrt{2}(\phi_c-\phi_s)}
   + {\rm h.c.} \}
   \;\; .
\end{equation}
%%%%%%%%%%%%%%%%%%%%%%%%%%%%%%%%%%%%%%%%%%%%%%%%%%%%%%%%%%%%%%%
In Eqs.\ (\ref{Umklapp}) -- (\ref{Peierls}) we have used 
$N_c = N_s = 0$
at half filling.
The parameter $a$
is a short-distance cutoff of the order of the lattice spacing, i.e.\ of order one.
The Luttinger parameters $g_{c,s}$  and the
charge and spin velocities $v_{c,s}$
can either be calculated perturbatively or 
from the Bethe ansatz solution of the Hubbard model
\cite{Lieb68,Schulz95}.
In the following we focus on the role of the Klein factors.
In the literature it is common practice either to ignore them
or to replace them by Majorana fermions \cite{Schulz00,Marston2002}.
It is argued that the latter approach -- which neglects
the number changing property of the Klein factors -- should be justified in the
thermodynamic limit.
In the following we aim to present a more rigorous approach.
Due to the conservation of charge and spin all combinations of
Klein factors appearing in the Hamiltonian (\ref{hamilt}) can be expressed
in terms of the operators $A_{\up} = F_{R\up}^+ F_{L\up}$ and
$A_{\dn} = F_{R\dn}^+ F_{L\dn}$ plus their hermitean conjugates.
In particular, the four-fermion terms arising from Umklapp and backscattering
read
%%%%%%%%%%%%%%%%%%%%%%%%%%%%%%%%%%%%%%%%%%%%%%%%%%%%%%%%%%%%%%%
\begin{eqnarray}
\label{fourterms}
F^+_{R\up} F^+_{R\dn} F_{L\dn} F_{L\up}
& = & F^+_{R\up} F_{L\up} F^+_{R\dn} F_{L\dn}
\;\; = \;\; A_{\up} A_{\dn} \;\; , \\
F^+_{R\up} F^+_{L\dn} F_{R\dn} F_{L\up}
& = & F^+_{R\up}  F_{L\up} F^+_{L\dn} F_{R\dn}
\;\; =\;\; A_{\up} A_{\dn}^+ \;\; .
\end{eqnarray}
%%%%%%%%%%%%%%%%%%%%%%%%%%%%%%%%%%%%%%%%%%%%%%%%%%%%%%%%%%%%%%%
Since the Klein factors are unitary, $F_{\alpha}^+F_{\alpha} = F_{\alpha}F_{\alpha}^+ = 1$,
it is easy to show that
%%%%%%%%%%%%%%%%%%%%%%%%%%%%%%%%%%%%%%%%%%%%%%%%%%%%%%%%%%%%%%%
\begin{equation}
[A_{\up},A_{\up}^+] = [A_{\dn},A_{\dn}^+] = 0 \;\; ,
\end{equation}
%%%%%%%%%%%%%%%%%%%%%%%%%%%%%%%%%%%%%%%%%%%%%%%%%%%%%%%%%%%%%%%
and we may choose a basis where $A_{\sigma}$ and $A_{\sigma}^+$ are both diagonal.
From $ A_{\up}^+ A_{\up} = A_{\dn}^+ A_{\dn} = 1$ one concludes that the eigenvalues
of $A_{\sigma}$ are pure phase factors, i.e.\
%%%%%%%%%%%%%%%%%%%%%%%%%%%%%%%%%%%%%%%%%%%%%%%%%%%%%%%%%%%%%%%
\begin{eqnarray}
A_{\up}   |k_{\up},k_{\dn}\rangle  & = & {\rm e}^{ik_{\up}}      |k_{\up},k_{\dn}\rangle \;\;\;\;\;\;\;\;\;\;\;
A_{\up}^+ |k_{\up},k_{\dn}\rangle \;\; = \;\;{\rm e}^{-ik_{\up}} |k_{\up},k_{\dn}\rangle\\
\label{eq20}
A_{\dn}   |k_{\up},k_{\dn}\rangle  & = & {\rm e}^{ik_{\dn}}      |k_{\up},k_{\dn}\rangle \;\;\;\;\;\;\;\;\;\;\;
A_{\dn}^+ |k_{\up},k_{\dn}\rangle \;\; = \;\; {\rm e}^{-ik_{\dn}}|k_{\up},k_{\dn}\rangle
\end{eqnarray}
%%%%%%%%%%%%%%%%%%%%%%%%%%%%%%%%%%%%%%%%%%%%%%%%%%%%%%%%%%%%%%%
with $0 \le k_{\sigma} < 2 \pi$.
The terms $\sim J_{c,s}^2$ appearing in $H_0$ do not commute with the Klein factors;
however it appears reasonable to neglect them in the
thermodynamic limit $L \rightarrow \infty$.
We will come back to this question in Sec.~\ref{Sec4}.
We thus replace the Klein factors in $H_1, H_2$ and $H_3$ by their eigenvalues,
and obtain
%%%%%%%%%%%%%%%%%%%%%%%%%%%%%%%%%%%%%%%%%%%%%%%%%%%%%%%%%%%%%%%
\begin{equation}
\label{Umklapp1}
H_1  = \tilde U \int_0^L dx \;
   \{ {\rm e}^{i(k_{\up}+k_{\dn})} {\rm e}^{-i 2\sqrt{2}\phi_c} + {\rm h.c.}\}
\end{equation}
\begin{equation}
\label{Backscattering1}
H_2 = \tilde U \int_0^L dx \;
   \{ {\rm e}^{i(k_{\up}-k_{\dn})} {\rm e}^{-i 2\sqrt{2}\phi_s} + {\rm h.c.}\}
\end{equation}
\begin{equation}
\label{Peierls1}
H_3 =  \tilde u \int_0^L dx \;  \{
   i {\rm e}^{ik_{\up}}{\rm e}^{-i\sqrt{2}(\phi_c+\phi_s)} +
   i {\rm e}^{ik_{\dn}}{\rm e}^{-i\sqrt{2}(\phi_c-\phi_s)} + {\rm h.c.}\} \;\; .
\end{equation}
%%%%%%%%%%%%%%%%%%%%%%%%%%%%%%%%%%%%%%%%%%%%%%%%%%%%%%%%%%%%%%%
As a result the Hamiltonian of the dimerized Hubbard model separates into
different sectors of purely bosonic Hamiltonians which are
labeled by $k_{\up}$ and $k_{\dn}$. Note that when replacing
Klein factors by Majorana
fermions \cite{Schulz00} one obtains only the eigenvalues $\pm i$ for
the two-fermion terms and $\pm 1$ for the four-fermion terms, i.e.\
continuity is lost.
Shifting the field operators according to $\phi_{c,s} \rightarrow \phi_{c,s}
+ (k_{\up} \pm  k_{\dn})/2\sqrt{2}$,
the phase factors can be absorbed, with the result 
%%%%%%%%%%%%%%%%%%%%%%%%%%%%%%%%%%%%%%%%%%%%%%%%%%%%%%%%%%%%%%%
\begin{equation}
\label{sine_Gordon}
H = H_0 + 2 \tilde U \int_0^L dx \;
       (\cos 2\sqrt{2}\phi_c + \cos 2\sqrt{2}\phi_s)
       + 4 \tilde u \int_0^L dx \;
        \sin \sqrt{2}\phi_c \cos \sqrt{2}\phi_s \;\; .
\end{equation}
%%%%%%%%%%%%%%%%%%%%%%%%%%%%%%%%%%%%%%%%%%%%%%%%%%%%%%%%%%%%%%%
In this sine-Gordon-like Hamiltonian the operator constraint
$[\phi_{c,s}]_{q=0}=0 $, see Eq.\ (\ref{varphi}),
has to be replaced by
%%%%%%%%%%%%%%%%%%%%%%%%%%%%%%%%%%%%%%%%%%%%%%%%%%%%%%%%%%%%%%%
\begin{equation}
\int_0^L dx \;  \phi_{c,s} = \frac{L}{2\sqrt{2}}(k_{\up} \pm  k_{\dn}) \;\; .
\end{equation}
%%%%%%%%%%%%%%%%%%%%%%%%%%%%%%%%%%%%%%%%%%%%%%%%%%%%%%%%%%%%%%%

\section{The self-consistent harmonic approximation}

In order to study the charge and spin gaps in the dimerized Hubbard model we use the
self-consistent harmonic approximation (SCHA) in which the exponentials of field operators
appearing in (\ref{Umklapp1}) -- (\ref{Peierls1}) are replaced by quadratic forms.
We introduce the trial Hamiltonian
%%%%%%%%%%%%%%%%%%%%%%%%%%%%%%%%%%%%%%%%%%%%%%%%%%%%%%%%%%%%%%%
\begin{equation}
\label{Htrial}
H_{\rm tr}  = \sum_{\alpha=c,s}\int_0^L
    \frac{dx}{2\pi}\left\{\frac{v_\alpha}{g_\alpha} (\partial_x\phi_\alpha)^2 +
      v_\alpha g_\alpha (\partial_x\theta_\alpha)^2 +
      \frac{\Delta_\alpha^2}{v_\alpha g_\alpha} \phi_\alpha^2 \right\}
\end{equation}
%%%%%%%%%%%%%%%%%%%%%%%%%%%%%%%%%%%%%%%%%%%%%%%%%%%%%%%%%%%%%%%
which provides us with a variational estimate for the ground state energy
%%%%%%%%%%%%%%%%%%%%%%%%%%%%%%%%%%%%%%%%%%%%%%%%%%%%%%%%%%%%%%%
\begin{eqnarray}
\label{varenergy}
\tilde E & = &
      \langle H_0 \rangle_{\rm tr} + \langle H_1 \rangle_{\rm tr}
     +\langle H_2 \rangle_{\rm tr} + \langle H_3 \rangle_{\rm tr}\; , \\
\frac{\tilde E}{L} & = & \frac{E_{\rm tr}}{L} - \sum_{\alpha=c,s}
\frac{\Delta_\alpha^2\langle\phi_\alpha^2 \rangle_{\rm tr}}{2\pi v_\alpha g_\alpha}
+ E(k_\up, k_\dn)
\; . \end{eqnarray}
%%%%%%%%%%%%%%%%%%%%%%%%%%%%%%%%%%%%%%%%%%%%%%%%%%%%%%%%%%%%%%%
Here the expectation value is with respect to the
ground state of $H_{\rm tr}$, $E_{\rm tr}$ is its ground state energy, and
%%%%%%%%%%%%%%%%%%%%%%%%%%%%%%%%%%%%%%%%%%%%%%%%%%%%%%%%%%%%%%%
\begin{equation} \label{eq26}
E(k_\up, k_\dn) = 2 B_1 \cos(k_\up + k_\dn) + 2 B_2 \cos(k_\up - k_\dn) - 2 B (\sin k_\up + \sin k_\dn )
\end{equation}
%%%%%%%%%%%%%%%%%%%%%%%%%%%%%%%%%%%%%%%%%%%%%%%%%%%%%%%%%%%%%%%
with
%%%%%%%%%%%%%%%%%%%%%%%%%%%%%%%%%%%%%%%%%%%%%%%%%%%%%%%%%%%%%%%
\begin{equation}
\label{Bdef}
B_1  =  \tilde U \; {\rm e}^{-4 \langle \phi^2_c \rangle_{\rm tr}}, \; \;
B_2  =  \tilde U \; {\rm e}^{-4 \langle \phi^2_s \rangle_{\rm tr}}, \; \; 
B  =  \tilde u \; {\rm e}^{- \langle \phi^2_c \rangle_{\rm tr}}
{\rm e}^{- \langle \phi^2_s \rangle_{\rm tr}}\; .
\end{equation}
%%%%%%%%%%%%%%%%%%%%%%%%%%%%%%%%%%%%%%%%%%%%%%%%%%%%%%%%%%%%%%%
Minimizing $\tilde E$ with respect to $\Delta_c$ and $\Delta_s$
yields the gap equations
%%%%%%%%%%%%%%%%%%%%%%%%%%%%%%%%%%%%%%%%%%%%%%%%%%%%%%%%%%%%%%%
\begin{eqnarray}
\label{gapequation1}
  \frac{\Delta_c^2}{2\pi v_c g_c } & = & - 4 B_1 \frac{\partial E_0}{\partial B_1}
 - B \frac{\partial E_0}{\partial B} \\
\label{gapequation2}
  \frac{\Delta_s^2}{2\pi v_s g_s } & = & - 4 B_2 \frac{\partial E_0}{\partial B_2}
 - B \frac{\partial E_0}{\partial B}
\end{eqnarray}
%%%%%%%%%%%%%%%%%%%%%%%%%%%%%%%%%%%%%%%%%%%%%%%%%%%%%%%%%%%%%%%
where $E_0$ is the minimum of
$E(k_\up, k_\dn)$ (see Table 1).
%%%%%%%%%%%%%%%%%%%%%%%%%%%%%%%%%%%%%%%%%%%%%%%%%%%%%%%%%%%%%%%
\begin{vchtable}
\begin{tabular}{|c|c|c|c|}
\hline
range & $k_{\up}$ & $k_{\dn}$ & $E_0(B_1,B_2,B)$ \\
\hline
 & & & \\
$0 < B < 2B_2$ & $\arcsin\frac{B}{2B_2}$ & $\pi - \arcsin\frac{B}{2B_2}$ &
$- 2 B_1 - 2 B_2 - \frac{B^2}{B_2}$ \\
 & & & \\
\hline
& & & \\
$2B_2 < B$ & $\frac{\pi}{2}$ & $\frac{\pi}{2}$ & $-4B - 2 B_1 + 2 B_2$ \\
 & & & \\
\hline
\end{tabular}
\caption{Minimum $E_0$ of $E(k_\up, k_\dn)$ -- see Eq.~(\ref{eq26}) --
which is used in the gap equations (\ref{gapequation1}) and (\ref{gapequation2}).
The first line is applicable for $u=0$ (Hubbard model without dimerization)
whereas the second line applies for $u > 0$.}
\end{vchtable}
%%%%%%%%%%%%%%%%%%%%%%%%%%%%%%%%%%%%%%%%%%%%%%%%%%%%%%%%%%%%%%%
In order to solve these equations analytically we consider the case $U > 0$
where $g_s = 1$ and $g_c < 1$, and restrict ourselves to the limit of small dimerization $u$.
Since $H_{\rm tr}$ is quadratic in the bosonic fields it is straightforward to calculate
%%%%%%%%%%%%%%%%%%%%%%%%%%%%%%%%%%%%%%%%%%%%%%%%%%%%%%%%%%%%%%%
\begin{eqnarray}
\label{phi2}
\langle \phi^2_c \rangle_{\rm tr} & = & \frac{g_c}{2} \ln\frac{\Delta_0}{\Delta_c} \\
\langle \phi^2_s \rangle_{\rm tr} & = & \frac{g_s}{2} \ln\frac{\Delta_0}{\Delta_s}
\end{eqnarray}
%%%%%%%%%%%%%%%%%%%%%%%%%%%%%%%%%%%%%%%%%%%%%%%%%%%%%%%%%%%%%%%
where $\Delta_0$ is a cutoff-dependent energy scale of the order of the bandwidth,
and $\Delta_{c,s} < \Delta_0$ is assumed.
For nonzero dimerization $u > 0$
the solutions of the gap equations lie in the range $B > 2B_2$; thus the
second line of Table 1 with $\partial E_0/\partial B_1 = -2, \partial E_0/\partial B_2 = 2$
and $\partial E_0/\partial B = -4$ has to be used
in Eqs.\ (\ref{gapequation1}), (\ref{gapequation2}).
For $u \rightarrow 0$ the spin gap vanishes while the charge gap approaches
a constant according to \cite{Otsuka97}
%%%%%%%%%%%%%%%%%%%%%%%%%%%%%%%%%%%%%%%%%%%%%%%%%%%%%%%%%%%%%%%
\begin{eqnarray}
\label{chargegap}
\Delta_c(u) - \Delta_c(0) & \propto & u^{4/3} \\
\label{spingap}
\Delta_s(u) &\propto & u^{2/3}
\end{eqnarray}
%%%%%%%%%%%%%%%%%%%%%%%%%%%%%%%%%%%%%%%%%%%%%%%%%%%%%%%%%%%%%%%
with cutoff-dependent prefactors. The exponent $2/3$ that characterizes the
spin gap is in accordance with
the corresponding exponent of the dimerized antiferromagnetic
Heisenberg chain up to a logarithmic correction in the prefactor \cite{Cross79,Uhrig96}.
Since the Heisenberg model corresponds to the $U \rightarrow \infty$ limit of
the half-filled Hubbard model, this indicates that as far as the exponent is concerned
the SCHA result (\ref{spingap})
is exact and persists even in the strong-coupling
regime $U/t \gg 1$.
For $u > u_{\rm co}$ the behavior of the gaps is changed to \cite{Schuster99}
%%%%%%%%%%%%%%%%%%%%%%%%%%%%%%%%%%%%%%%%%%%%%%%%%%%%%%%%%%%%%%%
\begin{eqnarray}
\label{gap1}
\Delta_c(u) \approx \Delta_s(u) & \propto & u^{2/(3-g_c)}
\end{eqnarray}
%%%%%%%%%%%%%%%%%%%%%%%%%%%%%%%%%%%%%%%%%%%%%%%%%%%%%%%%%%%%%%%
where the crossover value
$u_{\rm co}$ is defined by $\Delta_s(u_{\rm co}) = \Delta_c(0)$.
In Fig.\ 1 we show  $\Delta_c(u) - \Delta_c(0)$ and $\Delta_s(u)$
as a function of $u$ for $U/t = 2$ as obtained from the
numerical solution of the gap equations. For comparison, the analytical results
(\ref{chargegap}) and (\ref{spingap}) are also given.
%%%%%%%%%%%%%%%%%%%%%%%%%%%%%%%%%%%%%%%%%%%%%%%%%%%%%%%%%%%%%%%
\begin{vchfigure}[ht]
%\centerline
{\includegraphics[width=10.0cm,height=6.0cm]{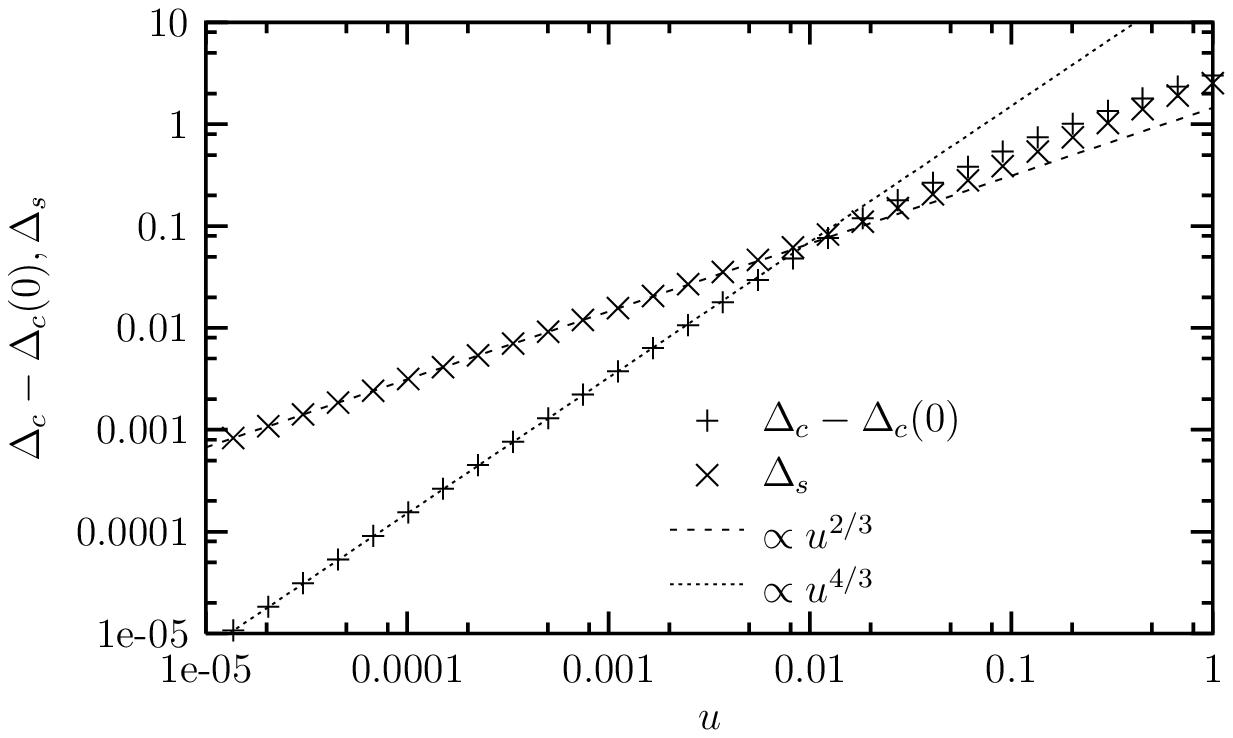}}
\caption{\label{fig1} Charge gap $\Delta_c$ and spin gap $\Delta_s$ (in units of $t$)
of the dimerized Hubbard model for $U/t = 2$ obtained within the SCHA.
[We used $v_{c,s}= v_F \sqrt{1 \pm U/\pi v_F}$, $g_s=1$, and $g_c=1/\sqrt{1+U/\pi v_F} $; note that $v_F = 2t $].
The straight lines are the analytic results (\ref{chargegap}) and (\ref{spingap})
valid for $u < u_{\rm co}$. Note that $\Delta_c(0)$ is subtracted from
the charge gap in  order to highlight the power law behavior.}
\end{vchfigure}
%%%%%%%%%%%%%%%%%%%%%%%%%%%%%%%%%%%%%%%%%%%%%%%%%%%%%%%%%%%%%%%
\section{Finite systems}
\label{Sec4}
For a finite system of length $L$ it is not possible to simply
replace the Klein factors by their eigenvalues
since the terms $\sim J_{s,c}^2$ in the Luttinger Hamiltonian (\ref{Luttinger})
do not commute with the $F's$.
However one may decouple the Klein factors from the bosonic fields using a variational
ansatz. To this end we introduce the ``Klein Hamiltonian"
%%%%%%%%%%%%%%%%%%%%%%%%%%%%%%%%%%%%%%%%%%%%%%%%%%%%%%%%%%%%%%%
\begin{eqnarray}
\label{HKlein}
H_{\rm tr}^B  & = & i B(F_{R\up}^+ F_{L\up} + F_{R\dn}^+ F_{L\dn})
+ B_1 F_{R\up}^+F_{R\dn}^+F_{L\dn}F_{L\up}
+ B_2 F_{R\up}^+F_{L\dn}^+F_{R\dn}F_{L\up} + {\rm h.c.} \nonumber \\
   &   &  + \frac{\pi}{4L}(v_c g_c J^2_c + v_s g_s J^2_s)
\end{eqnarray}
%%%%%%%%%%%%%%%%%%%%%%%%%%%%%%%%%%%%%%%%%%%%%%%%%%%%%%%%%%%%%%%
where $B_1, B_2$ and $B$ are variational parameters
to be determined self-consistently.
$H_{\rm tr}^B$ is of the form of a tight-binding Hamiltonian
for a particle moving on a $2d$ lattice in a harmonic potential.
A class of similar Hamiltonians for general potentials in $1d$
has been studied in \cite{Schoenhammer02}.
The choice of the trial Hamiltonian is equivalent to a decoupling of 
the non-linearities according to
%%%%%%%%%%%%%%%%%%%%%%%%%%%%%%%%%%%%%%%%%%%%%%%%%%%%%%%%%%%%%%%
\begin{eqnarray}
\label{meanfield}
&&{ F^+_{R\up}F^+_{L\dn}F_{R\dn}F_{L\up} {\rm e}^{-i 2\sqrt{2}\phi_s}
\rightarrow }  \cr
&&\langle F^+_{R\up}F^+_{L\dn}F_{R\dn}F_{L\up} \rangle_{\rm tr} {\rm e}^{-i 2\sqrt{2}\phi_s}
+ F^+_{R\up}F^+_{L\dn}F_{R\dn}F_{L\up} \langle {\rm e}^{-i 2\sqrt{2}\phi_s}\rangle_{\rm tr}
\; ,\end{eqnarray}
%%%%%%%%%%%%%%%%%%%%%%%%%%%%%%%%%%%%%%%%%%%%%%%%%%%%%%%%%%%%%%%
and analogously for the Umklapp and dimerization term.
This means that instead of
replacing the products of Klein factors by their {\it eigenvalues}
as in the thermodynamic limit, we now have to
replace them by their {\it expectation values} with respect to the
ground state of $H_{\rm tr}^B$. In both cases one ends up with
a sine-Gordon type model like the one explicitly given in Eq.\ (\ref{sine_Gordon}).
It is conceivable that for a finite-size system the eigenvalues are not identical to
the expectation values, i.e.\ different sine-Gordon models arise.

The question is then whether identical sine-Gordon models arise at least in the thermodynamic
limit.
We can answer this question within the framework of the SCHA.
Here the introduction of the ``Klein Hamiltonian" amounts 
to replacing the quantity
$E_0(B_1, B_2, B)$ which enters the gap equations (and which is
explicitly given in Table 1) by the ground state energy of $H_{\rm tr}^B$.
Apart from this modification Eqs.\ (\ref{Bdef}) -- (\ref{gapequation2}) remain unchanged.
Consider first a system with both a spin and a charge gap.
For large systems the parameters $B$, $B_1$, and $B_2$ become size-independent,
with the consequence that the kinetic energy in $H_{\rm tr}^B$ dominates 
the confining potential. The ground state energy of 
$H_{\rm tr}^B$ is then
given by the minimum of $E(k_\uparrow, k_\downarrow) $, see Eq.~(\ref{eq26}), and
expectation values of the Klein factors are equal to their eigenvalues.

In the absence of dimerization, i.e.\ for the ``pure" Hubbard model, we encounter a different situation since
the spin gap vanishes. For large system size
one finds $B =0$, $B_1 = {\rm const }$, and $B_2 \propto \tilde U (a/L)^2$.
Thus the confining potential in the spin sector
dominates the corresponding kinetic energy (proportional to $B_2$) in the thermodynamic limit.
Hence the ground state of $H_{\rm tr}^B$ is the eigenstate of the current operator
with $J_s=0 $, and 
the expectation value of the operator 
$F^+_{R\up}F^+_{L\dn}F_{R\dn}F_{L\up}$ in this state is zero, 
in contrast to the eigenvalues which are  $\exp(i k_\up -i k_\dn ) $, see Eqs.\ (\ref{fourterms}) -- (\ref{eq20}).

\section{Conclusions}
We studied the role of Klein factors for the bosonized Hamiltonian of the dimerized Hubbard model.
Since the Klein operators do not commute with the total
spin and charge currents $J_{c,s}$ they cannot acquire a definite value at the same time
as the current operators.

The ground state of the gapped system is a superposition of many spin and charge states.
In this situation it is justified to choose a fixed phase for the Klein operators. The 
bosonized Hamiltonian is then the conventional sine-Gordon-like Hamiltonian, cf.\ Eq.~(\ref{sine_Gordon}).
In an ungapped system where the non-linearities introduced by backscattering
are (marginally) irrelevant operators, the ground state has a well defined current.
In this case the phase of the Klein operators is undetermined and Eq.~(\ref{sine_Gordon}) has
no justification.

We have extended the self-consistent harmonic approximation 
in such a way that Klein factors can be handled systematically, and
we worked out the theory for the Hubbard model.
In a previous paper \cite{Mocanu04} we applied the SCHA to spinless
fermions with nearest-neighbor interaction and dimerization,
and studied in detail finite-size effects.
In the spinless case the Klein Hamiltonian $H_{\rm tr}^B$ can
be mapped onto a Mathieu equation which can be solved analytically
in certain limits.
It turns out that finite-size corrections to the gap equation
are not important as long as the dimerization gap is larger than the finite-size gap.
In addition, the finite-size formalism
allows to calculate the size-dependence of the
Drude weight in the gapped regime \cite{Mocanu04}.

\begin{acknowledgement}
We thank Kurt Sch\"onhammer and Cosima Schuster for helpful discussions.
Financial support from the Deutsche Forschungsgemeinschaft (SFB 484) is acknowledged.
\end{acknowledgement}


\begin{thebibliography}{10}
\bibitem{Tomonaga50}
S.\ Tomonaga, Prog.\ Theor.\ Phys.\ {\bf 5}, 544 (1950).

\bibitem{Haldane81}
F.~D.~M.\ Haldane, Phys.\ Rev.\ Lett. {\bf 47}, 1840 (1981).

\bibitem{Delft98}
J.~von~Delft and H.~Schoeller, Ann.~Phys.\ (Leipzig) {\bf 7}, 225 (1998).

\bibitem{Schulz00}
H.~J.~Schulz, G.~Cuniberti, and P.~Pieri, in {\it Field Theories for Low-Dimensional
Condensed Matter Systems: Spin Systems and Strongly Correlated Electrons},
edited by G.\ Morandi {\it et al.} (Springer, New York, 2000).

\bibitem{Coleman75}
S.~Coleman, Phys.~Rev.~D {\bf 11}, 2088 (1975).

\bibitem{Fukuyama85}
H.\ Fukuyama and H.\ Takayama, in {\em Electronic properties of
Inorganic One-Dimensional Compounds}, edited by P.\ Monceau (D.\ Reidel, Dordrecht,
Holland, 1985).

\bibitem{Gogolin93}
A.~Gogolin, Phys.\ Rev.\ Lett. {\bf 71}, 2995 (1993).

\bibitem{Mocanu04}
C.~Mocanu, M.~Dzierzawa, P.~Schwab, and U.~Eckern,
J.\ Phys.: Condens.\ Matter {\bf 16}, 6445 (2004).

\bibitem{Lieb68}
E.~H.~Lieb and F.~Y.~Wu, Phys.\ Rev.\ Lett. {\bf 70}, 2453 (1968).

\bibitem{Schulz95}
H.~J.\ Schulz, Int.\ J.\ Mod.\ Phys.\ B {\bf 5}, 57 (1991);
 in {\it Proceedings of Les Houches Summer School LXI},
edited by E.\ Ackermans, G.\ Montambaux, J.\ L.\ Pichard, and J.\ Zinn-Justin
(Elsevier, Amsterdam, 1995), p.\ 553.

\bibitem{Marston2002}J.~O.\ Fj{\ae}restad and J.~B.\ Marston, Phys.\ Rev.\ B {\bf 65}, 125106 (2002).

\bibitem{Otsuka97}H.\ Otsuka, Phys.\ Rev.\ B {\bf 56}, 15609 (1997).

\bibitem{Cross79}M.~C.\ Cross and D.~S.\ Fisher, 
   Phys.\ Rev.\  B {\bf 19}, 402 (1979).

\bibitem{Uhrig96}
G.~S.~Uhrig and H.~J.~Schulz,
Phys.\ Rev.\ B {\bf 54}, R9624 (1996).


\bibitem{Schuster99}
C.~Schuster, {\it Random and periodic lattice distortions in one-dimensional
Fermi and spin systems} (Shaker, Aachen, 1999) [PhD Thesis, Universit\"at Augsburg (1999)].

\bibitem{Schoenhammer02}
K.~Sch\"onhammer, Phys.\ Rev.\ A {\bf 66}, 014101 (2002).

\end{thebibliography}
\end{document}